\documentclass[prb,twocolumn,a4paper,showpacs,floatfix,superscriptaddress]{revtex4}  

\usepackage{graphicx}
\usepackage{amsmath}
\usepackage{amssymb}
\usepackage{amsthm}

\graphicspath{{./IMAGES/}}

\begin{document}
\title{Magnetic quantum dots and rings in two dimensions}

\author{C. A. Downing}
\email[]{downing@ipcms.unistra.fr}
\affiliation{Institut de Physique et Chimie des Mat\'{e}riaux de Strasbourg, Universit\'{e} de Strasbourg, CNRS UMR 7504, F-67034 Strasbourg, France}

\author{M. E. Portnoi}
\affiliation{School of Physics,
University of Exeter, Stocker Road, Exeter EX4 4QL, United Kingdom}
\affiliation{International Institute of Physics, Universidade Federal do Rio Grande do Norte, RN, Brazil}

\date{\today}

\begin{abstract}
We consider the motion of electrons confined to a two dimensional plane with an externally applied perpendicular inhomogeneous magnetic field, both with and without a Coulomb potential. We find that as long as the magnetic field is slowly-decaying, bound states in magnetic quantum dots are indeed possible. Several example cases of such magnetic quantum dots are considered in which one can find the eigenvalues and eigenfunctions in closed form, including two hitherto unknown quasi-exactly solvable models treated with confluent and biconfluent Heun polynomials. It is shown how a modulation of the strength of the magnetic field can exclude magnetic vortex-like states, rotating with a certain angular momenta and possessing a definite spin orientation, from forming. This indicates one may induce localization-delocalization transitions and suggests a mechanism for spin-separation.
\end{abstract}

\maketitle

\section{\label{intro}Introduction}

The energy levels of an electron in both a uniform magnetic field and harmonic oscillator potential (so-called Fock-Darwin levels) have been known since the 1920's.\cite{Fock, Darwin} When only the magnetic field is present, the states are the famous Landau levels.\cite{Landau, Dingle} These single particle exact solutions have formed the basis for more detailed research involving electrons in a magnetic field, for example in the Laughlin wavefunction\cite{Laughlin} and other many-body wavefunctions,\cite{Pfannkuche, Harju, Kainz} in the role of electron-electron interactions,\cite{Maksym} electron-phonon interactions\cite{Titeica, Apalkov} and excitons.\cite{Elliott, Lerner, anyonexciton} It is therefore worthwhile to investigate other magnetic profiles which admit analytic solutions.

There is a continued interest in two dimensional (2D) electron gases exposed to inhomogeneous magnetic fields,\cite{NogaretEPL, Xu} due to the variety of fundamental physics that can be explored, including the fractional quantum Hall effect, superconductivity and spintronics.\cite{Nogaret} Various methods exist for realizing such nonuniform fields, including the use of patterned gates made out of ferromagnets\cite{Ye, Henini, Lange} or superconductors.\cite{Bending, Geim92} Last year, an inhomogeneous magnetic field was created using a thin film of type-II superconducting niobium in close proximity to the studied system of graphene.\cite{Lindvall2015} Another method to probe a spatially varying magnetic field is to use molecular beam epitaxy regrowth technology to produce a non-planar 2D electron gas, and then to apply a constant magnetic field to the curved structure.\cite{Foden}

Theoretically, nonuniform magnetic fields have been extensively studied in magnetic quantum dots and magnetic barriers, both sharply defined\cite{Peeters1993, Matulis, Solimany, Chang, Ihm, Rosas, Reijniers2001, KimA, Sim, Kocsis, Masir} and smoothly defined.\cite{Muller, Ibrahim, Reijniers1999, Lee} There have also been many works on hydrogen atoms\cite{Parfitt, Collapse} in external magnetic fields, with both constant\cite{Halonen, Taut, Kravchenko} and nonhomogeneous\cite{Freire1, Freire2, LeeA} fields considered. In particular, the importance of spin effects have been examined in detail for magnetically confined excitons\cite{Freire2, Freire3}. Furthermore, progress on the many-body problem has been initiated with the treatment of two interacting electrons in a magnetic quantum dot.\cite{Mallon}

Here, we investigate several different inhomogeneous magnetic fields perpendicular to a 2D sheet of electrons. Contrary to previous works, we treat slowly decaying magnetic fields, dropping off as either $1/r$ or $1/r^{3/2}$, and include the effect of a cut-off at the origin. Notably, a $1/r$ magnetic field has recently been theoretically employed to trap electrons with a linear dispersion,\cite{Roy} which is hard to achieve with scalar potentials\cite{Downing, Stone, Churchill} but can be accomplished in various magnetic field configurations.\cite{Martino2007}

\begin{figure}[tbp]
 \includegraphics[width=0.45\textwidth]{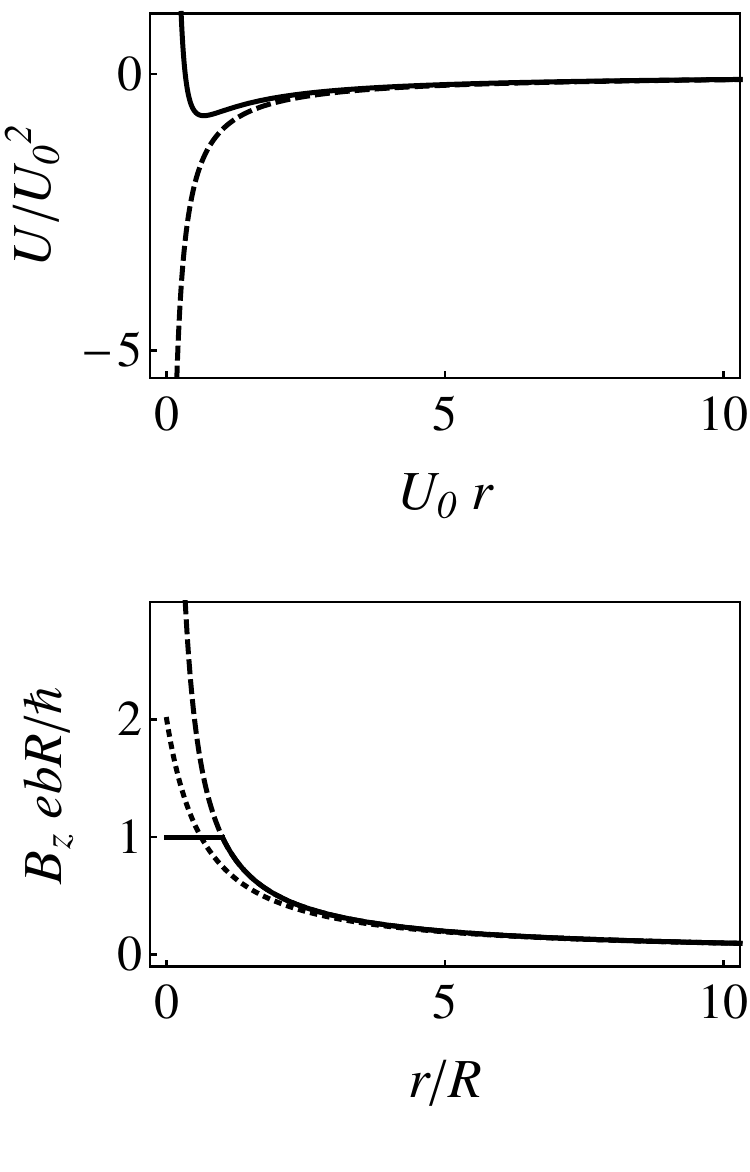}
 \caption{(Upper panel) A plot of the Coulomb potential (dashed line), as well as the modified Coulomb potential (solid line) considered in Eq.~\eqref{eq19}. Here $U_1 = 1/3$. (Lower panel) The inhomogeneous magnetic fields considered, without a cutoff at small distances (dashed line) as in Eq.~\eqref{eq10}, and with a smooth (dotted line) or sharp (solid line) regularization scheme, as in Eq.~\eqref{eq30} and Eq.~\eqref{eq21} respectively.}
 \label{fig1}
\end{figure}

Recently, it was shown that an electron cannot be trapped in a magnetic quantum dot defined by quickly decaying fields, due to an asymptotic reduction of the Schr\"{o}dinger equation to Bessel's differential equation.\cite{Masir} However, here we find that for slowly-decaying fields, dropping slower than $1/r^2$, there can indeed be square-integrable, truly confined states in magnetic traps. Our method of attacking this problem is via exact and quasi-exactly solvable models,\cite{Turbiner, DowningHeun, qes2} and we make use of the increasingly influential Heun functions.\cite{Ronveaux} Most notably, our integrable models unveil that the strength of the magnetic field determines whether certain rotations of the vortex-like states are prohibited or not, such that one may provoke localization-delocalization transitions by adjusting the field strength. As the nature of the exclusion of some vorticities (which have a set azimuthal quantum number) in the magnetic quantum dots is dependent on the electron spin orientation, it suggests that the system can act to spin polarize charged electronic modes. We also consider the consequences of introducing a Coulomb (or modified Coulomb) potential, please see Fig.~\ref{fig1}~(a) for all potential profile sketches.

The Schr\"{o}dinger-Pauli Hamiltonian, which captures the interaction between the spin of an electron and a magnetic field $\boldsymbol{B} = \nabla \times \boldsymbol{A}$ (entering via the minimal coupling substitution $\boldsymbol{p} \to \boldsymbol{p} + e \boldsymbol{A}$), is
\begin{equation}
\label{eq00}
 H = \frac{1}{2 M} \left( \boldsymbol {\sigma} \cdot \left[ \boldsymbol{p} + e \boldsymbol{A} \right] \right)^2,
\end{equation}
where $\boldsymbol {\sigma}$ are Pauli's spin matrices and $M$ is the electron mass. Eq.~\eqref{eq00} naturally arises from a Foldy-Wouthuysen transformation\cite{Foldy} of the Dirac equation in the non-relativistic limit, keeping the Zeeman term and neglecting all higher terms (Darwin term, Pauli spin-orbit coupling term, and so on). For a 2D planar system experiencing an electrostatic potential energy $V(r)$ and a perpendicular magnetic field $\boldsymbol{B} = (0, 0, B_z (r))$, the Pauli equation reads 
\begin{equation}
\label{eq01}
 \left( \frac{1}{2 M} \left[  \boldsymbol{p} + e \boldsymbol{A} \right]^2 + \tau \mu_B B_z \right) \Psi_{\tau} + V(r) \Psi_{\tau} = E \Psi_{\tau},
\end{equation}
where $\mu_B$ is the Bohr magneton, $\tau = \pm 1$ takes into account the electron spin orientation, and now the magnetic field is related to the vector potential by $B_z = r^{-1} \partial_r (r A_{\theta} (r))$. Eq.~\eqref{eq01} is separable in polar coordinates with the wavefunction $\Psi_{\tau}(r, \theta) = (2 \pi)^{-1/2} e^{i m \theta}\psi_{\tau}(r)$, where $m=0, \pm 1, \pm 2...$ is the azimuthal quantum number. In what follows, energy $E$ is rescaled via $\varepsilon = 2 M E/ \hbar^2$ and $U(r) = 2 M V(r)/ \hbar^2$. 

One can show that a (purely) magnetic quantum dot decaying asymptotically like $B_z \sim r^{-\gamma}$, where $\gamma$ is a positive number, only leads to square-integrable solutions if $0 < \gamma < 2$. Otherwise, the vector potential terms in Eq.~\eqref{eq01} drop out at large distances such that the solutions are Bessel functions. Now we present a plausibility argument as to why this precludes bound states: to be normalizable, this long range solution must be the modified Bessel function of the second kind, which imposes $\varepsilon < 0$. The magnetic quantum dot at short range must have (almost by definition) a region of approximately constant magnetic field, such that the solutions here look like Landau wavefunctions. Taking the limit of this short range region becoming large, the eigenvalues will tend to the (positive valued) Landau levels, which contradicts the original requirement of $\varepsilon < 0$.

The rest of this paper is devoted to solutions of Eq.~\eqref{eq01}, and is organized as follows. We study an electron in a Coulomb potential in an inhomogeneous magnetic field in Sec.~\ref{hydrogen}, and introduce a regularization scheme for the field in Sec.~\ref{hydrogenreg}. We introduce a quasi-exactly solvable model in Sec.~\ref{qes}, adding further weight to our arguments that confinement in magnetic quantum dots is possible. A free electron in a magnetic quantum dot is treated in Sec.~\ref{cutoff} and finally we draw some conclusions in Sec.~\ref{conc}. Appendix~\ref{appendAA} comments on spinless particles whilst Appendix~\ref{appendBB} details some complimentary results for electrons in a magnetic quantum ring.

\section{\label{hydrogen}Coulomb problem with an inhomogeneous magnetic field}

We consider an electron under the Coulomb potential $U(r) = -U_0 / r$. In terms of the fine structure constant $\alpha = e^2 / (4 \pi \epsilon_0 \hbar c) \approx 1/137$, it has a strength defined by the inverse length $U_0 =  2 \alpha M c / \hbar \approx (0.26 \AA)^{-1}$. Let us choose as an inhomogeneous magnetic field
\begin{equation}
\label{eq10}
 B_z (r) = \frac{\hbar}{e} \frac{1}{b r},
\end{equation}
where the length $b$ effectively parameterizes the strength of the field, as shown in Fig.~\ref{fig1}~(b). It follows from the radial part of Eq.~\eqref{eq01}, that after introducing a new variable $\xi = r/b$, we need to solve
\begin{equation}
\label{eq11}
 \psi'' +  \xi^{-1} \psi' + [ \varepsilon b^2 - 1 + (U_0 b - 2 m - \tau) \xi^{-1} - m^2 \xi^{-2} ] \psi = 0,
\end{equation}
where $'$ represents taking a derivative with respect to $\xi$. The solution required behaves for small $\xi$ proportional to $\xi^{\pm |m|}$, whilst in the regime $\varepsilon b^2 < 1$ it behaves asymptotically ($\xi \to \infty$) as $\psi \approx e^{\pm \kappa \xi}$, where
\begin{equation}
\label{eq12}
 \kappa = (1 - \varepsilon b^2)^{1/2}.
\end{equation}
Hence we seek a square-integrable solution in the form
\begin{equation}
\label{eq13}
	\psi = \tfrac{c}{b} \times \xi^{|m|} e^{- \kappa \xi} w(\xi),
\end{equation}
where $c$ is a normalization constant, such that we obtain for $w$ the equation
\begin{multline}
\label{eq14}
	\xi w'' + (b_0 - 2 \kappa \xi) w' - 2 \kappa a_0 w = 0, \\
	\text{where} \quad a_0 = \tfrac{1}{2} + |m| - \tfrac{U_0 b - 2 m - \tau}{ 2 \kappa}, \quad b_0 = 1 + 2|m|.
\end{multline}
This is a form of Kummer's differential equation, which has as a solution Kummer's function\cite{Gradshteyn}
\begin{equation}
\label{eq15}
	w = F(a, b, 2 \kappa \xi) = \sum\limits_{q=0}^\infty \frac{(a)_q}{(b)_q} \frac{(2 \kappa \xi)^q}{q !},
\end{equation}
where the Pochhammer symbol $(c)_q = \Gamma(c+q)/\Gamma(c)$ is defined in terms of the Gamma function $\Gamma(z)$. Terminating the power series to ensure normalizable solutions, we set $a = -n$, where $n$ is a negative integer (or zero). Consequently, we obtain the energy levels 
\begin{equation}
\label{eq16}
	\varepsilon_{n, m} b^2 = 1 - \left( \frac{U_0 b  - 2 m - \tau}{1 + 2 n + 2 |m|} \right)^2, \quad U_0 b > 2 m + \tau,
\end{equation}
which is dependent on both quantum numbers, $m$ and $n$. Notably, the $2 m$ dependence in Eq.~\eqref{eq16} is reminiscent of the degeneracy of zero-energy states bound by a Coulomb potential statically screened by a 2D electron gas.\cite{Gal} The confined state spectrum is bounded from above by $\varepsilon b^2 < 1$, a boundary at which there is an accumulation of highly oscillatory states with $n >> 1$. The infinite number of energy levels cease at some lower bound, which is dependent on the system parameters. The `s-state' $(n, m) = (0, 0)$ energy is given by $\varepsilon b^2 = U_0 b (2 \tau - U_0 b)$, which can be either positive or negative (or zero). Furthermore, the high vorticity (azimuthal quantum number) limit $(n, m) = (0, \pm \infty)$ unveils a collection of zero-energy states with $\varepsilon b^2 \simeq 0$.

\begin{figure}[tbp]
 \includegraphics[width=0.45\textwidth]{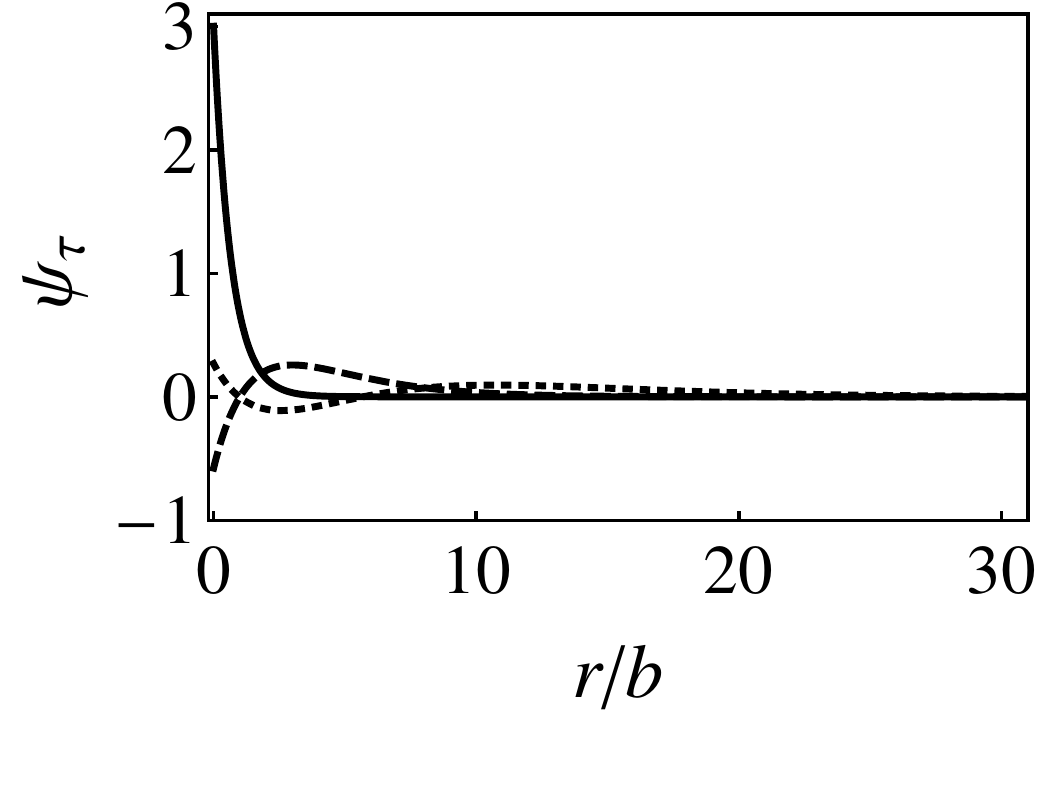}
 \caption{A plot of the radial part of the wavefunctions, $\psi_{\tau}(r)$,  for the spin-polarized $m = 0$, $\tau = -1$ bound states with $n = 0$ (solid line) $n = 1$ (dashed line) and $n = 2$ (dotted line). The signs of the wavefunctions were chosen so that the curves can be clearly distinguished.}
 \label{fig2}
\end{figure}

All bound state wavefunctions decay exponentially at large distances according to the localization length $\zeta = b/ \left(1-\varepsilon b^2\right)^{1/2}$. In Fig.~\ref{fig2}, we depict the radial part of the wavefunctions $\psi_\tau(r)$  for zero angular momentum states ($m=0$) for the lowest quantum numbers $n = 0, 1, 2$. The curves highlight the expected nodal structure and decay. In general, the corresponding radial probability distributions have a ring-like appearance, and the most probable radius increases as $n$ increases.

The condition $U_0 b > 2 m + \tau$ appearing in Eq.~\eqref{eq16} ensures that the value of the eigenenergy indeed terminates the special function Eq.~\eqref{eq15}, such that the size of $b$ tunes the allowable quantum states, each with a certain angular momentum and spin orientation. Namely, the states are rotating vortices described by a restricted azimuthal quantum number governed by $-\infty < m < m_{\tau}^{*}$, with the upper bound
\begin{equation}
\label{eq163434343}
 m_{\tau}^{*} = \bigg\lceil \frac{U_0 b  - \tau}{2} \bigg\rceil 
\end{equation}
given in terms of the ceiling function. Therefore one is able to induce successive localization-delocalization transitions in the system by continuously modulating the field strength, which ejects one-by-one bound states from the magnetic quantum dot which are no longer able to be supported as governed by the criterion Eq.~\eqref{eq163434343}. 

\begin{figure}[tbp]
 \includegraphics[width=0.5\textwidth]{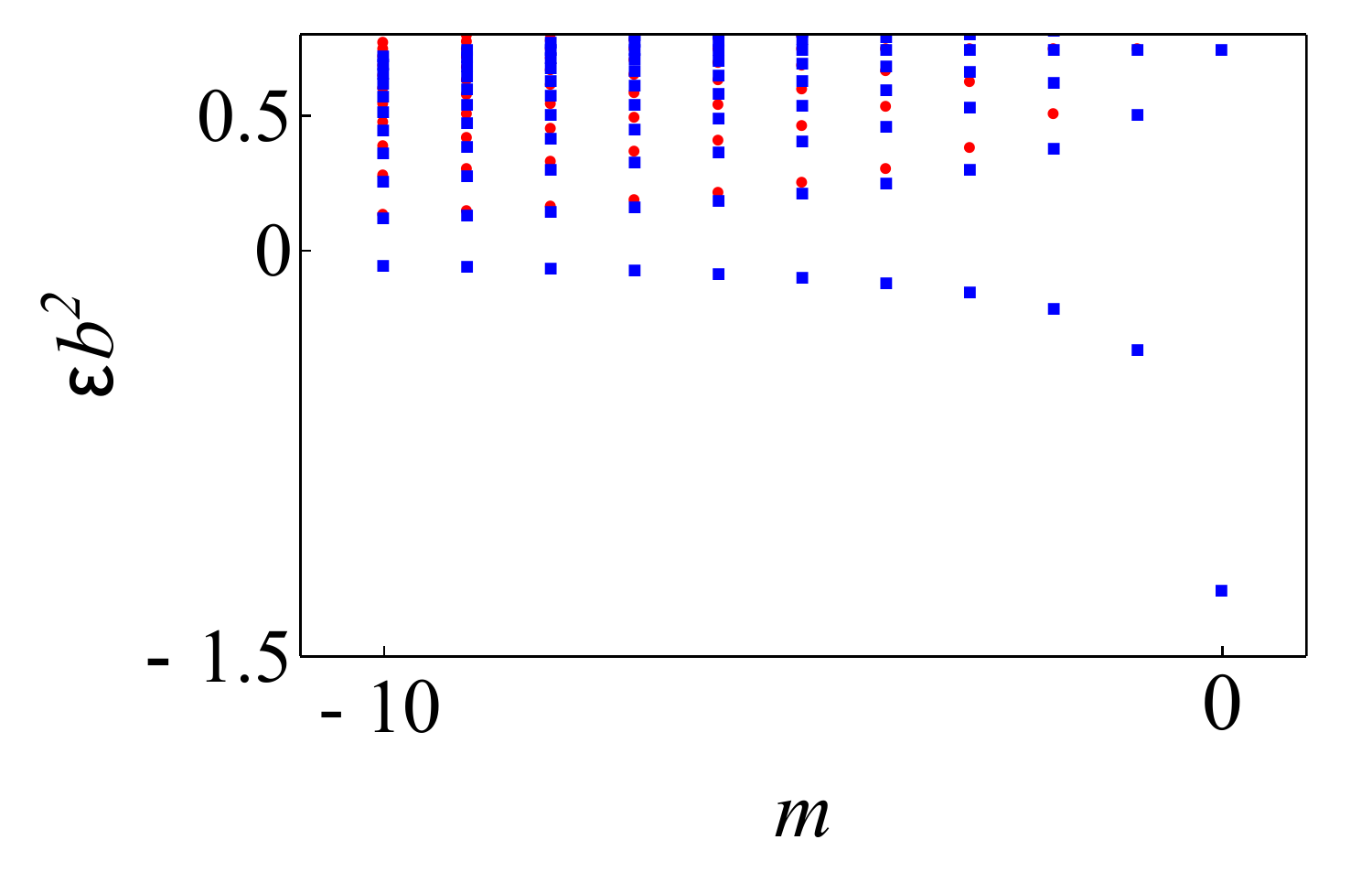}
 \caption{(Color online) A plot of the energy spectra as a function of angular momentum as given by Eq.~\eqref{eq16}, for both of the spin orientations $\tau = 1$ (red circles) and $\tau = -1$ (blue squares). Here $U_0 b = 1/2$.}
 \label{fig3}
\end{figure}

The nature of what states may or may not be trapped in the magnetic quantum dot gives rise to a mechanism of spin-polarization. For example, for a field strength $-1 < U_0 b \le 1$, it follows from Eq.~\eqref{eq163434343} that the $m=0$ states are only supported for the $\tau = -1$ spin orientation, and not for the $\tau = 1$ orientation. We plot in Fig.~\ref{fig3} the energy spectra as a function of angular momentum showing exactly this situation. Notably, the low-energy part of the spectrum is spin-polarized, with a large energy gap between the spin-separated eigenstates. The complete spin-polarization of the $m=0$ states holds for all $n$ and so potentially a large number of fermions per vorticity, raising the possibility of detecting such vortex states in magnetometery experiments.\cite{Usher} For higher energies (larger values of the quantum number $n$) the influence of the Zeeman term diminishes.  

\section{\label{hydrogenreg}Coulomb problem with a regularized magnetic field}

It is instructive to check whether the results of Sec.~\ref{hydrogen} are robust against the introduction of a regularization scheme as $r \to 0$. Accordingly, we introduce a new length scale $R$ and consider a magnetic field with the spatial dependence
\begin{equation}
\label{eq30}
 B_z (r) = \frac{\hbar}{e} \frac{1}{b R} \frac{2 + r/R}{(1 + r/R)^2},
\end{equation}
as displayed in Fig.~\ref{fig1}~(b). Now, the wavefunction must still behave when $r \sim 0$ like $\psi \approx r^{|m|}$ and the $r \to \infty$ behavior is also unchanged: $\psi \approx e^{\pm \kappa \xi}$, where $\kappa = (1-\varepsilon b^2)^{1/2}$ and $\xi = r/b$. We therefore assume a solution in the form
\begin{equation}
\label{eq31}
	\psi = \tfrac{c}{b} \times \xi^{|m|} e^{- \kappa \xi} w(\xi),
\end{equation}
which yields the rather unwieldy $2$nd-order equation
\begin{equation}
\label{eq32}
	\xi w'' + \left( \tfrac{1 + 2 |m|}{\xi} - 2 \kappa \right) w' 
	+ \left(  \tfrac{\Xi}{\xi} + \Upsilon  \right) w = 0.
\end{equation}
where the auxiliary parameter $\Xi$ and function $\Upsilon$ are given by
\begin{subequations}
\label{eq372}
\begin{align}
	\Xi = U_0 b - \kappa - 2 \kappa |m|,  \\
	\Upsilon = \frac{1 - \tfrac{2 \tau b}{R} + \tfrac{b \xi}{R} \left( 2 - \tfrac{\tau b}{R} \right) }{(1+\xi b / R)^2} - \frac{2 m b}{R} \frac{1}{1+\xi b / R}.
	\end{align}
\end{subequations}
Making the natural (and final) switch of the independent variable $\zeta = 1+ \xi b/R = 1 + r/R$, along with the following substitution
\begin{equation}
\label{eq38992}
	w = \zeta^{\tfrac{1}{2} - |\tfrac{R}{b} +\tfrac{\tau}{2}|} v(\zeta),
\end{equation}
brings the more convenient equation
\begin{equation}
\label{eq382}
	\xi v'' + \left( \alpha + \tfrac{\beta+1}{\zeta} + \tfrac{\gamma+1}{\zeta}  \right) v' 
      + \left( \tfrac{\mu}{\zeta} + \tfrac{\nu}{\zeta - 1}  \right) v = 0,
\end{equation}
which is the canonical form of the confluent Heun equation. In fact, a more useful parameterization is achieved with the help of the parameter transformation $(\mu, \nu) \to (\delta, \eta)$ via $\mu = \tfrac{1}{2} \left( \alpha - \beta - \gamma + \alpha \beta - \beta \gamma \right) - \eta$ and $\nu = \tfrac{1}{2} \left( \alpha + \beta + \gamma + \alpha \gamma + \beta \gamma \right) + \delta + \eta$. The reverse mapping is $\delta = \mu + \nu - \tfrac{\alpha}{2} \left( \beta + \gamma + 2 \right)$ and $\eta = \tfrac{\alpha}{2} \left( \beta + 1 \right) - \mu - \tfrac{1}{2} \left( \beta + \gamma + \beta \gamma \right)$. Hence, the parameters appearing in Eq.~\eqref{eq32} take a simple form
\begin{subequations}
\label{eq332}
\begin{align}
	\alpha = - 2 \kappa \tfrac{R}{b},~\beta = - | 2\tfrac{R}{b} + \tau|,~\gamma = 2 | m|,  \\
	 \delta = U_0 R + 2 \tfrac{R^2}{b^2} - \tfrac{R}{b} \left( 2 m + \tau \right) ,~\eta = \tfrac{1}{2} + \tfrac{R}{b} \left( 2 m + \tau \right) - 2 \tfrac{R^2}{b^2}.
	\end{align}
\end{subequations}
The local (Frobenius) solution built around the regular singular point $\zeta = 0$, with the radius of convergence $|\zeta| < 1$, is given by the confluent Heun function
\begin{equation}
\label{eq321}
	H_C(\alpha, \beta, \gamma, \delta, \eta, \zeta) = \sum\limits_{n=0}^\infty v_n(\alpha, \beta, \gamma, \delta, \eta, \zeta) \zeta^n,
\end{equation}
where the coefficients $v_n$ are given by the three-term recurrence relation\cite{Ronveaux}
\begin{equation}
\label{eq3211}
	A_n v_n = B_n v_{n-1} + C_n v_{n-2},
\end{equation}
which is subject to the initial conditions $v_{-1} = 0$, $v_{0} = 1$, where
\begin{subequations}
\label{eq32123}
 \begin{align}
  A_n &= 1 + \tfrac{\beta}{n}, \label{conda} \\
   \begin{split}
  B_n &= 1 + \tfrac{1}{n} \left( \beta + \gamma - \alpha - 1 \right)  \\
				&\quad{} + \tfrac{1}{n^2} \left( \eta - \tfrac{1}{2} (\beta + \gamma - \alpha) - \tfrac{\alpha \beta}{2} + \tfrac{\beta \gamma}{2} \right)  , \label{condb} \\
				 \end{split} \\
  C_n &= \tfrac{\alpha}{n^2} \left( \tfrac{\delta}{\alpha} + \tfrac{\beta + \gamma}{2} + n - 1 \right). 
 \end{align}
\end{subequations}
To obtain a bound state solution one needs to reduce the confluent Heun function to a confluent Heun polynomial of some degree $N$. Thus we need two successive terms in the three-term recurrence relation Eq.~\eqref{eq3211} to disappear, terminating the infinite power series appearing in Eq.~\eqref{eq321}. This requirement results in two termination conditions, which both need to be satisfied simultaneously. Thus the model is quasi-exactly solvable\cite{Turbiner} (QES) or explicitly solvable only in certain circumstances. Firstly, let us impose $C_{N+2} = 0$ or equivalently
\begin{equation}
\label{eq191dd}
	\frac{\delta}{\alpha} + \frac{\beta + \gamma}{2} + N + 1 = 0,
\end{equation}
re-arranging for the eigenvalues we obtain
\begin{equation}
\label{eq191}
	\varepsilon_{N, m}^{QES} b^2 = 1 -  \left( \frac{ \tfrac{1}{2} U_0 b + \tfrac{R}{b} - m - \tfrac{\tau}{2}}{1 + N + |m| - |\tfrac{R}{b} + \tfrac{\tau}{2}|} \right)^2.
\end{equation}
This expression is subject to the constraint of Eq.~\eqref{eq191dd} and therefore the ability to prohibit certain states from forming is robust against regularization of the magnetic quantum dot.

Secondly, let us force $v_{N+1} = 0$, such that it follows from Eq.~\eqref{eq3211} that all further terms in the series vanish identically - the function is now a polynomial of degree $N$. In practice this requires solving a polynomial in $\eta$. We illustrate this with the example of the $N=1$ state, with $U_0 R = 1$. Then for successively lower $m$, one obtains the solutions with $\tau = 1$: $\varepsilon_{1, 0} = -0.436$ with $b/R = 1.359$, $\varepsilon_{1, -1} = -0.007$ with $b/R = 1.840$ and $\varepsilon_{1, -2} = -0.000753$ with $b/R = 1.934$. The corresponding solutions when  $\tau = -1$ are: $\varepsilon_{1, 0} = -1.397$ with $b/R = 0.868$, $\varepsilon_{1, -1} = -0.509$ with $b/R = 0.961$ and $\varepsilon_{1, -2} = -0.328$ with $b/R = 0.981$. The existence of these solution refutes the belief there are no bound states in magnetic quantum dots. Notably, the case of a spinless particle in a regularized magnetic field can be treated much more simply, as is shown in Appendix~\ref{appendAA}.

It should be noted that the solutions presented here and in Sec.~\ref{hydrogen} are also valid in a more general electrostatic potential, also plotted in Fig.~\ref{fig1}~(a), given by 
\begin{equation}
\label{eq19}
	U(r) = -\frac{U_0}{r} + \frac{U_1^2}{r^2},
\end{equation}
which is also important, for example in systems with potentials decaying like the inverse square of distance\cite{DowningWhitt} such as a repulsive antidot confinement potential in a 2D electron gas\cite{Bogachek} and quantum rings.\cite{Tan} One only needs to make modifications in some places to the angular momentum quantum number in the solutions presented here. 

\section{\label{qes}The inverse $3/2$ magnetic model}

Remarkably, one can illustrate with another QES model a second counter-example of electron confinement in an inhomogeneous magnetic field decaying slower than $1/r^2$, with help from the exotic biconfluent Heun functions. In the model, we choose the field
\begin{equation}
\label{eq80}
 B_z (r) = \frac{\hbar}{e} \frac{1}{a^{1/2}} \frac{1}{r^{3/2}},
\end{equation}
where $a$ is a length scale marking the range of the field. The wavefunction should behave at short- and long-range like $\psi \approx r^{|m|}$ and $\psi \approx e^{-\sqrt{-\varepsilon}r}$ respectively, with $\varepsilon < 0$. Working in the variable $\xi = (2\sqrt{-\varepsilon}r)^{1/2}$, after one tries the ansatz
\begin{equation}
\label{eq81}
	\psi = \tfrac{c}{a} \times \xi^{2 |m|} e^{- \xi^2/2} w(\xi),
\end{equation}
to peel-off the asymptotics, one obtains a form of the so-called biconfluent Heun equation
\begin{equation}
\label{eq82}
	\xi w'' + \left( 1+\alpha  -2 \xi^2 \right) w'  
	- \left(  \tfrac{\delta}{2} + \left[ 2  + \alpha - \gamma \right] \xi  \right) w = 0.
\end{equation}
This equation has as a solution the biconfluent Heun function\cite{Ronveaux} 
\begin{equation}
\label{eq89}
	w = H_B\left(\alpha, \beta, \gamma, \delta, \xi\right) = \sum\limits_{n=0}^\infty \frac{v_n \left(\alpha, \beta, \gamma, \delta \right)}{(1+\alpha)_n} \frac{\xi^n}{n!},
\end{equation}
where the coefficients $v_n$ satisfy a three-term recurrence relation
\begin{equation}
\label{eq879}
	v_{n+2} = A_n v_{n+1} + B_n v_n,
\end{equation}
with $v_0 = 1$ and $v_1 = \tfrac{1}{2} \left( \delta + \beta (1 + \alpha) \right)$ where
\begin{subequations}
\label{eq8789}
\begin{align}
	A_n = (n+1)\beta + \tfrac{1}{2} \left( \delta + \beta (1+ \alpha) \right), \\
	B_n = (n+1)(n+1+\alpha )(2n +2 + \alpha - \gamma).
	\end{align}
\end{subequations}
In our case, the parameters in Eq.~\eqref{eq82} are 
\begin{equation}
\label{eq83}
	\alpha = 4 |m|, \quad \beta = 0, \quad \gamma = \frac{2 U_0 a - 8}{(-\varepsilon)^{1/2} a}, \quad \delta = \frac{4 \sqrt{2} (4 m + \tau)}{(-\varepsilon)^{1/4} a^{1/2}} .
\end{equation}
The biconfluent Heun function reduces to a polynomial when two conditions are met.\cite{Ronveaux} Firstly, $\gamma = 2 N + 2 + \alpha$, where $N$ is a positive integer, or equivalently
\begin{equation}
\label{eq85}
	\varepsilon_{N, m}^{QES} a^2 = -\left( \frac{U_0 a - 4}{1 + N + 2|m|} \right)^2, \quad U_0 a > 4. 
\end{equation}
Secondly, when Eq.~\eqref{eq85} holds the $(N+1)$-th coefficient in the series expansion is a polynomial in $\delta$ of order $N$. If $\delta$ is a root of that polynomial, then the $(N+1)$-th coefficient and indeed all subsequent coefficients $c_i$ are zero. The series has been truncated and $H_B(\alpha, \beta, \gamma, \delta, \xi)$ reduces to a biconfluent Heun polynomial $H_B = 1 + c_1 \xi + c_2 \xi^2 + ...$ of degree $N$.

In our case here, we need to solve $N$th-order polynomial equations for the remaining parameter $\varepsilon a^2$, which allows us to find closed form solutions in certain special cases. For example, let us consider the $m=1$ energy levels. When $N=1$, upon solving the resulting quadratic equation in $\delta$ and using Eq.~\eqref{eq85} one finds the energies $\varepsilon_{1, 1} a^2 = -400$ with $U_0 a = 84$~$(\tau = +1)$ and $\varepsilon_{1, 1} a^2 = -51.84$ with $U_0 a = 32.80$~$(\tau = -1)$. Similarly, for $N=2$ the closed form solutions arise for $\varepsilon_{2, 1} a^2 = -20.66$ with $U_0 a = 26.73$~$(\tau = +1)$ and $\varepsilon_{2, 1} a^2 = -2.68$ with $U_0 a = 12.18$~$(\tau = -1)$. For increasing $N$, a pattern arises of $(N-1)$ closed form solutions for each $\tau$. Thus we have found one more counterexample to the statement confinement in magnetic quantum dots is impossible, and in doing so have unveiled a novel toy model for the Pauli equation.

In fact, the limiting case of $m =0$ can be treated exactly when the Zeeman term is neglected $(\tau = 0)$ with the aid of a beautiful identity linking the biconfluent Heun and Kummer functions\cite{Ronveaux}
\begin{equation}
\label{eq86}
	H_B\left(\alpha, \beta, \gamma, \delta, \xi\right) = F \left( \tfrac{1}{2} + \tfrac{\alpha}{4} - \tfrac{\gamma}{4}, 1 + \tfrac{\alpha}{2}, \xi^2  \right),  
\end{equation}
\begin{equation*}
\label{eq87}
 \beta = \delta = 0, \quad \alpha \ne -n, \quad n = 0, 1...
\end{equation*}
Now, after terminating the infinite series of the Kummer function, one readily obtains the eigenspectra
\begin{equation}
\label{eq88}
	\varepsilon_{n, 0} a^2 = -\left( \frac{U_0 a - 4}{1 + 2 n} \right)^2, \quad U_0 a > 4,
\end{equation}
which again explicitly shows the characteristic feature of a threshold value of $U_0 a$ that must be obtained before bound states may form. A notable distinction, compared to the s-state solution Eq.~\eqref{eq16} for the $1/r$ decaying field, is that the eigenvalues are always negative.

\section{\label{cutoff}Electron in a magnetic quantum dot}

We now turn to a regularized inhomogeneous magnetic field for a free electron, which allows us to probe all states exactly, in the form of a magnetic dot
\begin{equation}
\label{eq21}
	B_z (r) = \frac{\hbar}{e} \frac{1}{b} \begin{cases} R^{-1}, \quad r \le R, \quad \text{(region I)} \\
r^{-1}, \quad r > R. \quad \text{(region II)} \end{cases}
\end{equation}
as displayed in Fig.~\ref{fig1}~(b). In region I, one can write down the solution in a constant magnetic field\cite{LandauLifshitz} as follows
\begin{equation}
\label{eq22}
	\psi_{I} = \tfrac{c_{I}}{b} \xi_{I}^{|m|/2} e^{-\xi_I/2} F(a_{I}, b_{I}, \xi_{I}),
\end{equation}
\begin{equation*}
\label{eq23}
	a_{I} = \tfrac{1}{2} ( 1 + |m| + m - \varepsilon b R + \tau), \quad b_{I} = 1 + |m|,
\end{equation*}
where the radial coordinate has been eliminated via $\xi_{I} = r^2/(2bR)$ and $c_{I}$ is a normalization constant. In region II, guided by the solution in Sec.~\ref{hydrogen}, one finds the solution
\begin{equation}
\label{eq24}
	\psi_{II} = \tfrac{c_{II}}{b} \xi_{II}^{|m|} e^{-\xi_{II}/2} U(a_{II}, b_{II}, \xi_{II}),
\end{equation}
\begin{equation*}
\label{eq25}
	a_{II} = \tfrac{1}{2} + \tfrac{2 m + \tau}{2 \kappa} + |m|, \quad b_{II} = 1 + 2 |m|,
\end{equation*}
in the variable $\xi_{II} = 2 \kappa r/b$. The Tricomi function, the second linearly independent solution to Kummer's differential equation, is defined by\cite{Gradshteyn}
\begin{multline}
\label{eq26}
	U(a, b, \xi) = \frac{\Gamma(1-b)}{\Gamma(a-b+1)} F(a, b, \xi) \\
		+ \frac{\Gamma(b-1)}{\Gamma(a)} \xi^{1-b} F(a-b+1, 2-b, \xi) ,
\end{multline}
which has the asymptotic behavior $U(a, b, \xi) \sim \xi^{-a}$ as $\xi \to \infty$. We discard Kummer's function as a physical solution due to its large $\xi$ expansion $F(a, b, \xi) \sim \xi^{a-b} e^{\xi}$. Imposing the boundary conditions of continuity of the wavefunction and its first spatial derivative at the boundary $R$, yields the constraint
\begin{equation}
\label{eq27}
	\frac{c_{II}}{c_{I}} = \left( \frac{b}{8 \kappa^2 R} \right)^{\tfrac{|m|}{2}} e^{\frac{R}{b}\left( \kappa - \tfrac{1}{4} \right)} \frac{F(a_{I}, b_{I}, \tfrac{R}{2 b} )}{U(a_{II}, b_{II}, \tfrac{2 \kappa R}{b})},
\end{equation}
and the following transcendental equation for determination of the eigenvalues
\begin{multline}
\label{eq28}
	\frac{a_{I}}{b_{I}} \frac{F(a_{I}+1, b_{I}+1, \tfrac{R}{2 b} )}{F(a_{I}, b_{I}, \tfrac{R}{2 b} )} + 2 \kappa a_{II} \frac{U(a_{II}+1, b_{II}+1, \tfrac{2 \kappa R}{b} )}{U(a_{II}, b_{II}, \tfrac{2 \kappa R}{ b} )} \\
+ \kappa - \frac{1}{2} = 0.
\end{multline}
This rich equation~\eqref{eq28} recovers the expected results in the limit of constant magnetic field, $B_z (r) = \hbar / e b R$, and the appropriate inhomogeneous magnetic field, $B_z (r) = \hbar / e b r$, respectively:
\begin{subequations}
\label{eq29}
 \begin{align}
  \varepsilon_{n, m} b^2 = \tfrac{b}{R} (1 + |m| + m + 2 n + \tau),~R/b >> 1,   \label{coupled1} \\
  \varepsilon_{n, m} b^2 = 1 - \left( \frac{2 m + \tau}{1 + 2 n + 2 |m|} \right)^2,~2 m + \tau < 0,~ R/b << 1. \label{coupled2}
 \end{align}
\end{subequations}
One notices how a modulation of the magnetic field effects the key dimensionless parameter $R/b$, such that in strong fields $R/b << 1$ one can exclude all states rotating with a positive angular momentum due to the requirement $2 m + \tau < 0$. In weak fields $R/b >> 1$ one recovers the celebrated Landau levels. States with different electron spin orientations are not treated symmetrically, as is seen from the condition in Eq.~\eqref{coupled2}. This implies magnetic vortex states trapped in magnetic quantum dots as a potential system to observe polarization of the electron species, since the confinement of a state with a certain spin orientation $\tau$ does not imply the partner state (with the same quantum numbers but with opposite spin orientation $-\tau$) is also confined. Additionally, as varying the magnetic field strength leads to successive vortex states undergoing localization-delocalization transitions as confined states are lost into the continuum, the setup is a plausible candidate for the basis of a magnetic storage device. In Appendix~\ref{appendBB} we derive results for electrons in a magnetic quantum ring, which leads to analogous conclusions. 

We should also mention that the addition of a magnetic flux tube to the problem leads to an extra phase factor in the wavefunction, accounted for by the replacement $m \to \tilde{m} = m + f$ (where $f$ is the number of flux quanta) such that one can now modulate this key physical parameter. The freedom of $\tilde{m}$ to take values away from purely integers also means that this setup requires a proper treatment including the von Neumann theory of self-adjoint extensions.\cite{Gupta}

\section{\label{conc}Conclusion}

Whilst it is true that confinement is not possible for 2D massive electrons in magnetic quantum dots defined by short-range magnetic fields, this is not the case for slowly-decaying magnetic fields. We have studied the counter-examples of fields decaying like $1/r$ or $1/r^{3/2}$, showing how the electrons can be trapped in quantized energy levels depending on two quantum numbers, the spin orientation and two parameters of the field, defining its strength and spatial extent. Interestingly, manipulation of the magnetic field strength allows one to exclude certain magnetic vortex states from forming, raising the possibility of both observing successive localization-delocalization transitions and spin polarization effects. We hope that this proposal stimulates further experimental work on trapping electrons with inhomogeneous magnetic fields in 2D systems.

\section*{Acknowledgments}

We acknowledge financial support from the CNRS through the PICS program (Contract No. 6384 APAG) and from the ANR under Grant No. ANR-14-CE26-0005 Q-MetaMat, as well as the EU H2020 RISE project CoExAN (Grant No. H2020-644076), EU FP7 ITN NOTEDEV (Grant No. FP7-607521), and the FP7 IRSES projects CANTOR (Grant No. FP7-612285), QOCaN (Grant No. FP7-316432), and InterNoM (Grant No. FP7-612624). We would like to thank V.~A.~Saroka for fruitful discussions and R.~Plant and J.~Page for a critical reading of the manuscript. 

\begin{appendix}

\section{\label{appendAA}Particle in a regularized magnetic quantum dot}

The case of a spinless particle (or arguably a particle in a fixed eigenstate of spin\cite{Gurtler, Hestenes}) follows by taking $\tau = 0$ in the main part of this paper. It also allows for an exact treatment of a particle in a regularized magnetic quantum ring, defined by
\begin{equation}
\label{eq345450}
 B_z (r) = \frac{\hbar}{e} \frac{1}{b} \frac{1}{\sqrt{R^2 + r^2}}.
\end{equation}
Now the wavefunction must behave when $r \sim 0$ like $\psi \approx \xi^{\pm \sqrt{|m|^2+R^2/b^2}}$, with $\xi = r/b$. Thus, with comparison to Sec.~\ref{hydrogen}, we notice that the effect of the regularization is to ensure that all eigenfunctions, including the s-state with $m=0$, have a ring-like structure. Remarkably, one can find the s-state eigensolution analytically via a formal analogy with Eq.~\eqref{eq11}, leading to the (unnormalized) solution
\begin{equation}
\label{eq31b}
 \psi = \xi^{\tfrac{R}{b}} e^{- \kappa \xi} F\left(-n, 1+2 \tfrac{R}{b}, 2 \kappa \xi\right), \quad n=0,1,2,...
\end{equation}
with the eigenenergy spectrum
\begin{equation}
\label{eq32b}
 \varepsilon_{n, 0} b^2 = 1 - \left( \frac{U_0 b}{1 + 2 n + 2 R/b} \right)^2.
\end{equation}
Of course the spectrum reduces in the limit $R/b << 1$, where the cutoff is of negligible importance, to the s-state solution of Eq.~\eqref{eq16}. Crucially, this result shows the analysis carried out in Sec.~\ref{hydrogenreg} is not misleading, in spite of a divergence in the field at the origin, and can be safely used as a toy model with small cutoffs $R/b << 1$.

Neglecting any electrostatic potential $(U_0 = 0)$, the s-state eigensolution Eq.~\eqref{eq31b} and Eq.~\eqref{eq32b} do not give us much useful information. In this limiting case, the wavefunction instead takes the form of a modified Bessel function of the second kind 
\begin{equation}
\label{eq37b}
 \psi = K_{R/b}\left( \kappa \xi\right), 
\end{equation}
in order to decay asymptotically at long range, which it does like $\psi \sim (\xi \kappa)^{-1/2} e^{-\xi \kappa}$. The quantization of the $m =0$ energy level is removed, all that is required is the inequality $\varepsilon b^2 < 1$ holds. However, the requirement of a square integrable wavefunction places the additional constraint $R/b < 1$, due to the singular nature of the wavefunction at short-range
\begin{equation}
\label{eq38b}
 \psi \sim (\xi \kappa)^{-R/b}, \quad \xi \to 0.
\end{equation}
Similar singular wavefunctions are well known in 2D, both in anyonic physics\cite{Grundberg} and in the scattering of Dirac fermions by cosmic strings.\cite{Gerbert, Hagen} The complete removal of a cut-off $R = 0$ leads to a logarithmic singularity at the origin, and as such forbids the $m = 0$ state.

One may gain further insight into this problem via an approximate analytical solution of the Schr\"{o}dinger equation with Eq.~\eqref{eq345450} and now $m \neq 0$. Introducing the replacement $(1+r^2/R^2)^{1/2} \approx (1+r/R)$, to ensure the correct behavior both at $r=0$ and $r>>R$ for the term which appears as the crossed term in Eq.~\eqref{eq01}, leads to the approximate eigenvalue expression 
\begin{equation}
\label{eq39b}
  \varepsilon_{n, m}^{app} b^2 \simeq 1 - \left( \frac{ U_0 b - 2 m }{1 + 2 n + 2 |m + R/b| } \right)^2, \quad U_0 b > 2 m,
\end{equation}
which is valid for strong fields $R/b << 1$. One notices that the effect of the cutoff is to increase the magnitude of the energy, compared to Eq.~\eqref{eq16}. A free spinless particle again exhibits the feature of removing most states with positive angular momenta, as found previously for electrons. 

\section{\label{appendBB}Electron in a magnetic quantum ring}

For completeness, we consider an analogous situation to Sec.~\ref{cutoff}, but now with a magnetic ring defined by
\begin{equation}
\label{eq991}
 B_z (r) = \frac{\hbar}{e} \frac{1}{b r} \Theta(r-R),
\end{equation}
where $\Theta(x)$ is the Heaviside step function. One can write down the solution using our knowledge from Sec.~\ref{cutoff}. Now for $r \le R$ (region $I$) the wavefunction simply becomes a Bessel function of the first kind $\psi_{I} = \tfrac{c_{I}}{b} J_{|m|} (\varepsilon^{1/2} r)$, whilst it is unchanged from Eq.~\eqref{eq24} when $r > R$ (region $II$). This analysis leads to a new transcendental equation to be solved for bound states
\begin{equation}
\label{eq99}
	2 a_{II} \frac{U(a_{II}+1, b_{II}+1, \tfrac{2 \kappa R}{ b})}{U(a_{II}, b_{II}, \tfrac{2 \kappa R}{ b})} 
		- \frac{\varepsilon^{1/2} b}{\kappa} \frac{J_{|m|+1} (\varepsilon^{1/2} R)}{J_{|m|} (\varepsilon^{1/2} R)}  + 1 = 0.
\end{equation}
This expression reduces in the limit of $R/b << 1$ to the problem of Sec.~\ref{hydrogen}, with the spectrum of Eq.~\eqref{coupled2}. Therefore, the regime of $R/b << 1$ again displays the criticality on the quantum number $m$.

\end{appendix}

\end{document}